
\magnification=1200
\baselineskip =22pt
\def\iras{{\it IRAS }}
\def\dmm{$\delta M/ M_{L}$}
\def\dmmg{$\delta M/ M_{g}$}
\def\etl{{\it et al.} }                               
\def\etal,{{\it et al.,}}                               

\def\ppp{\par \smallskip \noindent \hangindent .5in \hangafter 1}
\def\pt{ \ \ \ \ .}
\def\com{\ \ \ \ ,}

\def\a{({2 / 9 \pi} )^{1/3} }
\def\h1{$h^{-1}$}

\def\ap{\approx}
\def\H0{H_0}
\def\kms{{\rm km/s} }
\def\dcrit{$\delta_{crit} $ }
\def\frac#1#2{ {{#1}\over{#2}}}
\chardef\letter=11
\chardef\other=12
\def\unlock{\catcode`\@=\letter}%
\def\lock{\catcode`\@=\other}%
\unlock
\def\lsim{\mathrel{\mathpalette\@versim<}}
\def\gsim{\mathrel{\mathpalette\@versim>}}
\def\@versim#1#2{\lower0.2ex\vbox{\baselineskip\z@skip\lineskip\z@skip
  \lineskiplimit\z@\ialign{$\m@th#1\hfil##\hfil$\crcr#2\crcr\sim\crcr}}}
\def\VEV#1{\left\langle{#1}\right\rangle}

\lock

\centerline{\bf  LIMITS ON  THE PRIMORDIAL FLUCTUATION SPECTRUM}
\centerline{\bf  VOID SIZES AND CMBR ANISOTROPY}
\bigskip
\bigskip
\centerline{\bf T. Piran$^{1,2}$, M. Lecar$^{1}$, D. S. Goldwirth$^{1}$,
L. Nicolaci da Costa$^{1,3,5}$,  and G. R. Blumenthal$^4$}

\bigskip\bigskip
\centerline{\bf ABSTRACT}
\baselineskip = 19pt

We suggest to use the common appearance of voids with a scale of $5000
\kms$ in the galaxy distribution to estimate the power spectrum on
this scale.  We use a simple model for a gravitational formation of
voids and we compare the results with the matter fluctuations as
constrained by the CMBR observations of COBE.  We find that a power
spectrum $P(k)\propto k^n$ with $n \approx 1.25$ is compatible both
with COBE and with gravitational growth of large voids in an
$\Omega=1$ universe.  A Harrison-Zel'dovich spectrum, $n=1$,
normalized to produce the observed CMBR fluctuations, does not have
enough power for gravitational growth of voids with a diameter of $5000
\kms$ in an $\Omega=1$ Universe.  Such a spectrum would be compatible
if (i) void diameters are smaller ($3500 \kms$), the voids are
shallower or if the voids are rare (we assume that the universe is
void filled); (ii) $\Omega < 1$ i.e.  the Universe is open; (iii)
galaxies do not trace matter on very large scale or if (iv) the voids
do not grow gravitationally.

\bigskip\bigskip
\noindent
{\bf Subject Headings:} Cosmology -- Galaxies: clustering -- Cosmic
background radiation
\bigskip

\item {1.}
Harvard-Smithsonian Center for Astrophysics, Cambridge MA 02138, USA

\item {2.}
Racah Institute for Physics, The Hebrew University, Jerusalem, 91904 Israel.

\item {3.}
CNPq/Observatorio Nacional, Rio de Janeiro, Brasil.

\item {4.}
UCO/Lick Observatory, Board of Studies in Astronomy \& Astrophysics,
University of California, Santa Cruz, California 95064, USA.

\item {5.}
John Simon Guggenheim Fellow.

\vfill\eject
\baselineskip = 22pt
\noindent
{\bf I. INTRODUCTION}

Recent observations of fluctuations in the cosmic microwave background
radiation (CMBR) have provided strong support for the gravitational
instability theory of structure formation. Moreover, since temperature
fluctuations on scales larger than $1^o$ are expected to arise from
fluctuations in the potential field at the recombination era (Sachs
and Wolfe 1967), these measurements can be used to directly constrain
the primordial power--spectrum of matter fluctuations.  On very large
scales ($\approx$ 1000\h1 Mpc), the CMBR anisotropy detected by the
COBE DMR experiment (Smoot \etal, 1992) can be used to normalize the
power--spectrum of matter fluctuations.  On smaller scales, more
sensitive upper limits on the CMBR anisotropy have recently been
reported by Gaier \etal, (1992) based on the University of California
Santa Barbara (UCSB) South Pole degree-scale experiment while Meyer
(1993) has recently reported a detection on a comparable scale.  This
observation provides additional information on scales of about 100\h1
Mpc that combined with the COBE data can also be used to constrain the
shape of the power--spectrum for long wavelengths.

These recent results of CMBR experiments have had a considerable
impact on theories of structure formation.  Until recently, the
standard biased CDM model ($b = 1/\sigma_8 \approx 2.5$) involving cold
dark matter, Gaussian fluctuations and a Harrison-Zel'dovich spectrum
had been one of the most successful models for the formation of
structures in the universe.  The CDM scenario was very appealing
because of its simplicity, predictive power and the enormous success
it had in accounting for a vast number of observations ranging from
galaxy scales to about 10 \h1 Mpc.  On larger scales, the model was
less successful, having difficulties with the growing body of evidence
indicating the need for extra amounts of large-scale power (Davis
\etal, 1992).  The amplitude of the two--point correlation at large
angular scales for the APM sample (Maddox \etal, 1990a,b), the amplitude
of the variance in galaxy counts determined from deep redshift surveys
of \iras galaxies (Efstathiou \etal, 1990, Saunders \etal, 1992), the
large coherent motions detected on scales of about 50\h1 Mpc or larger
(Bertschinger \etal, 1990, Courteau 1992), and the existence (and the
frequency) of large voids in the galaxy distribution were some of the
observations requiring more power on large scales.

The consequences of these observations and the new CMBR measurements
to theories of large-scale structure formation are currently under
investigation. For instance, Efstathiou, Bond and White (1992) find
that normalizing the canonical CDM power--spectrum to the COBE
measurements requires $\sigma_8 \ap 1$ and $b \ap 1$ and leads to
conflicts with other observational data on small scales.  This has
motivated several authors to investigate different cosmological
scenarios including the tilted CDM model (Cen \etal, 1992), the mixed
C+HDM model (Davis \etal, 1992, Klypin \etal, 1992) and low-$\Omega$
CDM models with non-zero cosmological constant (Efstathiou, Bond and
White 1992, Kofman \etal, 1992), all trying to reconcile the small
and large scale properties, most based on the clustering properties of
galaxies.

An alternative and more general approach to the problem is to consider
a generic model--independent power-spectrum and examine the nature of
the constraints imposed by combining the CMBR observations on
large-scales, which are expected to be independent of the details
relating galaxies to the matter density field, with observations on
smaller scales.  For instance, if large-scale structures grow
gravitationally the amplitude of motions and the characteristics of
voids will be sensitive to the fluctuations of the {\it total} density
field and thus can be used to constrain the power-spectrum of the {\it
matter} fluctuations on scales of the order of 50\h1 Mpc.  Since the
process of galaxy formation is rather complex and models of the
hydrodynamical and star formation processes are still very preliminary
(Cen and Ostriker 1992a,b), it seems that properties that depend on
matter fluctuations are more suitable to constrain the primordial
power-spectrum than using clustering properties of {\it galaxies}.

An example of this approach is the analysis of Gorski (1992) who
investigated the bounds imposed by the new UCSB measurements to the
deviations from a Hubble flow.
In this paper we use a similar approach to investigate the implications
to the nature of the primordial fluctuation spectrum, imposed by
combining the CMBR measurements with the existence of large-scale
voids.
We
suggest here a simple quantitative model that estimates the
power required for gravitational formation of voids.
Towards this end we extend the earlier work of Blumenthal
\etal, 1991 (paper I) who examined the conditions for the
gravitational formation of voids from negative rms fluctuations in the
primordial power-spectrum. Our approach is similar in nature to that
of Gorski (1992) in the sense that our treatment is independent of the
details of any specific structure formation model.  Our primary goal
is to develop the formalism and investigate how the observed
properties of voids (size, underdensity and frequency) can be combined
with the CMBR data to constrain the power-spectrum of matter
fluctuations.  In section 2 we review the observational background.
The basic assumptions of our model and the condition for the
gravitational formation of a void at a given scale today are
reviewed in section 3.
In section 4 we show how the CMBR data is used to normalize a power--law
spectrum, while in section 5 we investigate the limits on the size of voids
imposed by the CMBR measurements on large and small scales.
A summary of our main conclusions is presented in section 6.

\bigskip\bigskip
\noindent
{\bf II. OBSERVATIONAL BACKGROUND}
\bigskip
Recent observations of the galaxy distribution strongly suggest that
galaxies tend to lie on wall-like features which bound large empty
regions, forming a closely packed network of voids (Kirshner \etal,
1981, 1983; De Lapparent, Geller \& Huchra 1986; Geller \& Huchra,
1989; da Costa \etal, 1988).  This property seems to be generic since
it is present in all the slices in the northern and southern
hemisphere that have been completed by the ongoing CfA2 and SSRS2
surveys (Geller \& Huchra 1989, da Costa 1991).  These surveys, which
densely probe the galaxy distribution out to about 15000 \kms, also
reveal that 5000 \kms voids are quite common. In fact, evidence that
5000 \kms may perhaps be a characteristic scale of the present--day
large-scale structure has recently been provided by the redshift maps
of the deep survey being conducted by Schectman \etl (1992). Combined
these surveys are qualitatively suggestive of a void-filled Universe,
with a characteristic scale of 5000 \kms. Below we investigate the
consequences of this interpretation.

It is important to emphasize that no suitable algorithm is currently
available to quantitavely study the properties of voids such as
abundance, size distribution, filling factor and density profile,
which depend on the amplitude of the large-scale density fluctuations
and on the dynamical evolution of primordial low--density regions.
Only some gross estimates such as typical size and underdensity can be
inferred from the existing redshift surveys, which suggest a large
abundance of voids ranging from 2500 to 5000 \kms in diameter
(Kirshner, 1993) with a typical underdensity estimated at about 20\%
of the mean (see Dey, Strauss \& Huchra , 1990, for a discussion the
Bo\"otes void).

Based on the qualitative evidence presented above, in this paper we
adopt the picture that the Universe is volume filled by 5000 \kms
voids with 20\% underdensity. Although there is still some debate
whether this is a fair description of the observed galaxy distribution
(it is possible that a typical size is 2500 \kms and that 5000
\kms are rare) and the underdensity is just a rough estimate
based on the galaxies we feel that it is of interest to explore the
theoretical implications of this picture.  In sections V and VI we
discuss how these various observational uncertainties affect our
conclusions.

Finally, we point out that
we interpret the common existence of these large voids
as a consequence of non-linear effects in the evolution
of the underlying matter distribution, which must also affect
the galaxy distribution since the observed voids are delineated
by galaxies.  This may be the origin of the bump observed
by Vogeley \etl (1992) in the power-spectrum of the galaxy distribution
at a characteristic scale of 5000 \kms.

\bigskip
\noindent
{\bf III. GRAVITATIONAL GROWTH OF VOIDS}
\bigskip

Following paper I, we focus our attention on the evolution of negative
amplitude fluctuations to explain the formation of the
observed voids in the galaxy
distribution.  We argue that the gravitational evolution of negative
fluctuations offer a natural explanation for the observed geometry of
the large-scale structure as they grow in size with time while
positive fluctuations of the same scale collapse.  The essence of our
model is that the present--day voids result from the gravitational
growth of small primordial negative fluctuations reaching shell
crossing today and that the requirement that voids grow
gravitationally can, in principle, yield a quantitative constraint on
the primordial spectrum of fluctuation on the scale of the observed
voids.

Ideally, a full {\it N}-body simulation with a large dynamical range
would be required to determine the amplitude of the primordial
fluctuations needed to produce {\it gravitationally} the observed
voids. Since suitable simulations are not currently available, (see
however, first attempts towards this direction by Van de Weygaert
(1991) and Van de Weygaert \& Van Kampen (1993)) here we follow paper I
and consider the growth of an isolated spherical inverted top-hat
void.  Despite the simplicity of the adopted model, its main features
are supported by recent N-body experiments of simple configurations of
several interacting voids carried out by Dubinski \etl (1993). In
these experiments it was found that the condition for formation of an
isolated void holds approximately, even for highly non--spherical
perturbations and when other negative or positive perturbations are
present.  These numerical simulations also reveal that voids at
shell--crossing are the most prominent being delineated by
high-density contrast walls. Only tenuous traces of smaller scale
voids, past the shell-crossing phase, are seen producing a
distribution that roughly resembles that actually observed for the
galaxies.

The required amplitude for the matter fluctuations
at horizon crossing to form a void at the present epoch
in a flat universe ($\Omega = 1$) was derived in paper I.  We assume
that at some initial time $t_i$ (e.g. at horizon crossing) there is an
inverted top-hat density distribution; i.e., for $r \le R_i$, $\rho =
\rho_c ( 1 - \epsilon_i)$ where $\rho_c$ is the critical density and
$\epsilon_i = 1 - \Omega_i$, and for $r > R_i$, $\rho = \rho_c$.
Initially this perturbation grows linearly. During this phase the
depth of the void changes with time while its size grows only with the
expansion of the universe. Eventually the perturbation becomes
non-linear and its size increases faster than the expansion of the
universe.  Finally at a time $t_{sc}$ given by $\epsilon_i
(t_{sc}/t_i)^{2/3} \approx \delta_{crit}$, (i.e. when $\epsilon_i
(1+z_i)/(1+z_{sc}) \approx \delta_{crit}$) shell crossing begins
(with the shell just inside the boundary being the first to cross). We define
the instant of shell crossing as the moment at which the void forms.
This definition might seems somewhat arbitrary.  However, at $t_{sc}$
the co-moving radius $R_{sc}$ has grown and it is 1.7 times the
initial comoving radius $R_{i}$. The density in the void is then
$1/1.7^3 \approx 0.2$ of the average density, which is the frequently
quoted value for the galaxy underdensity in the observed voids.  The
requirement for shell crossing may perhaps be too extreme, although it
naturally leads to underdensity values which are roughly comparable to
observations. Relaxing this condition and requiring the
underdensity to be 30\% of the mean (which corresponds to an expansion
over the comoving scale by a factor of 1.45) implies that the
amplitude of the initial perturbation could be decreased by a factor of
1.5.

We considered, in paper I, pure density fluctuations for which \dcrit =
4.5.  However, if the perturbation cross the horizon in a pure growing
mode the density perturbation will be accompanied by a velocity
perturbation and for such a perturbation \dcrit = 2.7, as pointed out
by Dubinski \etl (1993).  It is not clear what is the exact nature of
the perturbations as they enter the horizon. We decided to adopt here
the common growing mode assumption, keeping in mind the corresponding
uncertainty in \dcrit.

Two other properties are critical to our model.  First, the largest
voids at any epoch are those reaching shell crossing at that epoch,
since after shell-crossing voids grow relatively slowly.  As discussed
in paper I, this is satisfied for a wide range of power-spectra.
Second,  Dubinski \etl (1993) argue that if the fluctuations are
Gaussian-distributed then when rms negative fluctuations on a given
scale reach  shell crossing the corresponding voids will also occupy a
large fraction of the volume of the universe, resembling a void-filled
universe. We use this fact to identify $\epsilon$
with $\sigma$, the rms mass fluctuation on the scale of voids.  It
might be that in a more realistic model $1.5 \sigma$ or $2 \sigma$
fluctuations will be sufficient to produce a void-filled universe.
This introduces an additional uncertainty in our model which
is difficult to estimate (see section VI).

The condition for shell crossing implies an initial amplitude
$\epsilon_i = \delta_{crit} (1+z_{sc})/(1+z_i) $.  If we take $z_{sc}$
= 0 and $z_{i} = z_{rec}$ we have a condition for the amplitude of the
mass fluctuations at  recombination, $\epsilon_{rec}$.  These
fluctuations are also responsible for the CMBR fluctuations measured
today.  Therefore for a given form of the power-spectrum we can use
the observed fluctuations of the CMBR on large scales to normalize the
power-spectrum and determine the comoving scale that satisfies the
condition  $ \epsilon_{rec} = \delta_{crit}/(1+z_{rec}) $.  This would
then provide the size of  voids today, consistent with the CMBR
fluctuations.

Instead of working at the recombination era we can determine the
comoving scale of voids in terms of the {\it linear} power spectrum
today, $P_L(k) =  (1 + z_{rec})^2 P_{rec} (k) $.  We notice that
$\epsilon_i (1+z_i)/(1+z_{sc} ) $ can also be viewed as the amplitude
of the perturbation $\delta M /M_L$ had it continued to grow {\it
linearly} until today.  Since we are interested in the rms {\it linear}
fluctuation
in the density field $\sqrt{ <(\delta M /M)^2_L>}$ we can express it
in terms of $P_L(k)$ as:
$$
\VEV{\left(\frac{\delta M }{M}\right)_L^2 (R) } =
\frac{1}{2\pi^2} \int_0^\infty dk\; k^2 P_L(k)
\;{\cal W}^2 (kR) \com  \eqno(1)
$$
where ${\cal W} (kR)$ is a window function.  Using this we can
directly relate the appearance of voids to the {\it linear} amplitude
of the matter perturbations measured by COBE via the Sachs-Wolfe
effect, in exactly the same way as it was done in paper I.

For the window function ${\cal W}(x)$ in equation (1) we use the
Fourier transform of a Gaussian  which has been normalized to have the
same spatial volume as a top hat of radius $R$:
$$
{\cal W}(x) =\exp[ - {1\over 2} x^2 \a ] \pt \eqno(2)
$$
This choice of a window function assures the convergence of the
integrals in Eq. (1).

Before turning to the comparison between the value of  $\delta M/M_L$
needed for the formation of voids and the limits imposed by the CMBR
measurements it is worthwhile to emphasize that voids undergo shell
crossing well after the perturbation has become non-linear. At such
late stages of development the actual power spectrum bears no
resemblance to the  linear power spectrum,
$P_L(k)$.  The scale of an underdense region, once the evolution
becomes non-linear, increases in comoving coordinates by a factor of
1.7 at shell crossing, and an overdense ridge (formally, a caustic)
forms at the edge of the shell.  The linear power spectrum, $P_L(k)$,
which we use in Eq. (1) should not be confused with neither the
present non-linear $P(k)$ for the matter distribution nor with the
observed $P_g(k)$, the power--spectrum of the galaxy distribution
today.  Both will be modified by the highly non-linear evolution (see
e.g. Couchman \& Carlberg 1992). In addition, $P_g(k)$ will also be
affected by biasing factors (reflecting the relationship between the
distribution of galaxies and  matter) determined by dissipational
processes which are difficult to model.

On the other hand, we note that although the relation between $P_g(k)$
and $P(k)$ is undetermined in our model, we expect that the former
will show some signature of the non-linear evolution since the
galaxies do delineate the observed voids.  One possibility  would be
the presence of a peak in $P_g(k)$ on the  scale of the
voids.  It is noteworthy that Vogeley \etl (1992) detect a
feature near the 5000 \kms in the power-spectrum of the galaxy
distribution, which may be the signature of the characteristic scale
of the voids.  We point out, however,
that it is not clear how well defined this signature
should be. As an experiment we have computed the power-spectrum for
a Voronoi tesselation, which vaguely resembles the observed galaxy
distribution.  We find that although the power--spectrum
peaks at the characteristic length of the
cells, the peak is broad and may not be discernible in a more
realistic distribution with more power on small scales.

Similarly, $(\delta M/M)_L$ should not be confused with the expected
$\delta M/M $ for the matter distribution today.  In particular, we
emphasize that we cannot compare it to the recent measurements of
$\delta M/M_{g} $ by Efstathiou \etl (1990).  In fact, very non-linear
models such as the Voronoi tesselation yield values
of $\delta M/M$ which are small ($<$ 1), indicating that the
statistic $\delta M/M$ is not a suitable indicator of the dynamical
phase of the underlying matter distribution.
In conclusion, the existing measurements of $P_g(k)$ and \dmmg\quad
cannot rule out the existence of excess large-scale power and that
non-linear effects maybe taking place beyond 8\h1 Mpc.

\bigskip\bigskip
\noindent
{\bf IV. LIMITS ON THE POWER-SPECTRUM FROM CMBR DATA}
\bigskip
Smoot \etl (1992) describe the results of the DMR experiment on the
COBE satellite. The experiment has a beam width of $3.5^o$ and the
data extends from $\approx 5^o$ to $180^o$.  DMR measures the
temperature correlation function, $C_{obs} (\theta) = \langle \Delta T
(0) \Delta T (\theta) \rangle $ of the CMBR. Smoot \etl (1992) present
the  observed quadrupole moment: $C_{2~(obs)} = 13  \pm 4 \mu K $ of
the CMBR fluctuations and a  correlation function $\tilde C$ from
which the monopole, the dipole and the quadrupole terms were removed.
The tilde denotes the subtraction of those moments.

The fluctuations in the CMBR on  scales larger than one  degree, which
are of interest here, are  dominated by the Sachs-Wolfe effect (Sachs
\& Wolfe, 1967).  In a flat ($\Omega = 1$) universe, $C(\theta)$ can
be expressed in terms of the current {\it linear} power spectrum,
$P_L(k)$, as:
$$
{ C(\theta) \over T^2 } = \frac{ {H_o}^4 }{8\pi^2 c^4} \int_0^\infty
\frac{dk}{k^2} P_L(k) \frac{ \sin k\zeta}{k\zeta} \com
\eqno(3)
$$
(see e.g. Gorski, 1991). The angular dependence of $C(\theta) $ is
expressed through
$$
\zeta \equiv 2 R_h \sin ( \theta /2) \com \eqno(4)
$$
where $R_h=2c/H_o$ is the horizon size.

After removing the monopole, dipole and
quadrupole moments from $C(\theta)$ we obtain:
$$ {\tilde C(\theta) \over T^2 } = \frac{ {H_o}^4 }{8\pi^2 c^4}
\int_0^\infty \frac{dk}{k^2} P_L(k) \left\{ \frac{ \sin k\zeta}{k\zeta}
- \left( \frac{ \sin kR_h}{kR_h} \right)^2 - \frac{3\cos\theta
}{(kR_h)^2} \left( \frac{\sin kR_h}{kR_h} - \cos kR_h \right)^2
\right.
$$
$$
 - 5 \left( {3 \over 2} \cos^2\theta - {1 \over 2} \right)
\left[ \frac{3}{(k R_h)^2}
- \left( \frac{ \sin kR_h}{kR_h} \right)^2 \left( 2 + \frac{3}{(k R_h)^2}
\right) \right.
\eqno(5)
$$
$$
\left. \left.
- \frac{3}{(kR_h)^2} \left( {\sin kR_h \over kR_h} -\cos kR_h \right)^2
\left( 1 - \frac{3}{(k R_h)^2} \right) \right]
\right\}
$$

To estimate $\tilde C(\theta)$ we choose a power spectrum, $P_L(k)$, of
the form $P_L(k) = A_n k^n$.  This might be justified since we are
interested in extrapolation only from the scale of COBE measurements
to the scale of the voids.
For each $n$ we define  $F (\theta)$ such
that:
$$
{C(\theta) \over T^2} \equiv \frac { 2}{ \pi^2}
\left ( \frac{ {H_o} }{ 2  c } \right )^{n+3} A_n F(\theta)
\pt \eqno(6)
$$
The ratio $\VEV{ {\tilde C_{obs} / T^2 \tilde F }} $, averaged over
$\theta$, yields the amplitude $A_n H_o^{n+3} /(2^{n+2} \pi^2 c^{n+3})
$.  Explicitly, we use $\VEV{ {\tilde C_{obs} / T^2 \tilde F }} =
0.5[\tilde C_{obs}(10^o)/ T^2 \tilde F (10^o) + \tilde C_{obs} (20^o)/
T^2 \tilde F (20^o)]$ to obtain the results presented in Table I.
These results do not change significantly if we average over different
values of $\theta$ in the same range.  An alternative way to estimate
this ratio is to compare the quadrupole moment of the function $F$,
which we denote by $F_{2}$, with the observed quadrupole moment
measured by COBE, $C_{2~(obs)} = 13 \mu K $.  The resulting ratio is
also shown in Table I. It is slightly lower than $\VEV{ {\tilde
C_{obs} / T^2 \tilde F }} $, in agreement with the conclusions of
Smoot \etl (1992) and Wright \etl (1992) that the quadrupole moment is
somewhat lower than expected from $\tilde C(\theta)$. Note that there
is a better agreement for $n$ at the middle of the allowed range.
Again, in agreement with Smoot \etl (1992) and Wright \etl (1992), a
comparison of the analytical curves and the observed data show that $
0.6 \le n \le 1.6$.
As a final note we should mention that recently Gould (1993)
has argued that a proper analysis of the errors in the
estimates of the quadrupole moment suggest that the best fit
to the spectral index is $n=1.5$ rather than $n=1.1$ found
by Smoot \etl (1992).

The COBE data constrains the power-spectrum on very large scales.
On smaller scales
Gaier \etl (1992) recently reported a 95 \% confidence level upper
limit of $\Delta T/T$ for a Gaussian model on a scale of $1.2^o$ of
$1.4 \times 10^{-5}$.  If we extrapolate the $C(\theta)$
to this angular scale
(using  the amplitude given in Table I from the fit to
COBE)  we find $\sqrt{C(1.2^o)}=2.4 \times
10^{-5}$ for $n=1.5$ and $\sqrt{C(1.2^o)}=1.6 \times 10^{-5}$ for
$n=1$.  These values suggest that $n \le 1$.  However, Gaier \etl
(1992) describe their results in terms of $\Delta T/T$ for a Gaussian
model which is not directly comparable to $C(\theta)$ (Blumenthal
\etal, 1992; Gorski, 1992) and consequently
it is not clear how one should interpret this
constraint. In an attempt to compare our fit to COBE with Gaier \etl
results we present in Table I the values for $C_{1.2^o}(0)$,
where $C_{1.2^o}$ is convolved with a FWHM beam of $1.2^{o}$
(Mart\'{\i}nez-Gonz\'alez \& Sanz. 1989).
We find that for $n=1$ $\sqrt{C_{1.2}(0^o)}= 1.4 \times 10^{-5}$.
If this is the correct interpretation the constraint is even more
severe if
we recall that on the scale of $1^o$ additional effects such as peculiar
velocities (Doppler effect) contribute to $\Delta T/T$, while our $C(\theta)$
includes only the Sachs--Wolfe effect.

A more direct constrain on small scales has recently become available
from the work of
Meyer (1993), who report a detection of $\Delta T/T = 1.4 \pm 0.5 \times
10^{-5}$ for an experiment with a FWHM of $3.5^o$. Similar analysis of
our predicted values (see Table I) shows that in this case $n=1$
agrees well with this measurement. In fact,
the entire range of power-indexes  considered
$n=0.5$ to $n=1.5$ is within the quoted error bars of this experiment.

\bigskip\bigskip
\noindent {\bf V. VOIDS AND CMBR ANISOTROPY}
\bigskip

Using the results of the preceding sections we can now examine the
size of typical voids today which are consistent
with the constrains on the power-spectrum imposed by the CMBR
experiments.  Using the COBE data alone we find that
for a given power-law spectrum, we can integrate equation (3),
determine the appropriate normalization constant $A_n$, and
integrate equation (1) to obtain an estimate of $\delta M/M_{L}$ as
a function of the power index $n$:

$$
{\delta M \over M}_{L} (v) = \sqrt{  2^n
\left ( { 9 \pi  \over 2} \right)^{(n+3)/6}
\Gamma  \left( {n+3 \over 2} \right) }
\sqrt{ \VEV{ {\tilde C_{obs} \over T^2 \tilde F }}  }
\left ({ c \over  v } \right )^{(n+3)/2} \pt
\eqno(7)
$$
Here we have used the velocity, $v = H_0 R $, instead of the radius,
$R$, to obtain a formula that is independent of the Hubble constant.
We also remind the reader that
since at shell crossing  $R_{sc} \approx 1.7 R_{i}$,
a typical void with a diameter of $2 R_{sc} $ today
corresponds to an initial perturbation
with a comoving radius of $R_{i}$.

The condition imposed by the existence of voids that evolve
gravitationally from rms fluctuations can be expressed by equating
equation (7) to the value of \dcrit discussed in section III.
We adopt in the following the value \dcrit = 2.7 but remind  the reader
that \dcrit could be as low as 1.8 (if we consider underdensity of
30\% in the voids or if $1.5\sigma$ rather than $1\sigma$ fluctuations
produce the typical voids) or as high as 4.5 (if we adopt a pure density
initial perturbations).

Our results are summarized in figure 1, where we show the variation of
$\delta M/M_L$ as a function of $v_f = 3.4 \times v$, where $v$
corresponds to the comoving size in \kms of the unevolved primordial
perturbation.
We also represent the likely range of power
required to form voids, \dcrit.  A summary of our results is also
presented in Table I in columns (4) and (5) where we list for each $n$
the value of $\delta M /M_{L}$ for $v = 1500 km/s$, (which can lead to
the formation of 5000 \kms voids) and the scale at which the void
formation condition $\delta M/M_L = 2.7$ is satisfied, respectively.

Inspection of figure 1 shows that if we assume that the most likely value for
\dcrit is 2.7 than in order to form typical voids
of 5000 \kms in diameter we must consider a power-spectrum with
$n \approx 1.25$.  This could be the case if the
primordial spectrum is steeper than expected from inflation.
Since  there is no physical reason to expect an amplification of a
primordial scale--independent spectrum somewhere between the scale of
measurements of COBE and the scale of the  voids a
spectrum that steepens at short wavelengths would have had to be imprinted
in the initial conditions.

These results suggest that a void-filled distribution with
characteristic scale of 5000 \kms while it agrees with the
results of COBE and Meyer (1993) may already be in conflict with the
upper limits of the CMBR anisotropy at the $1^o$ scale. However, it is
unclear how the Gaier \etl (1992) limits apply since, as we
discussed earlier, these results are described in terms of $\Delta
T/T$ for a Gaussian model for the correlation function which is not
directly comparable to the computed $C(\theta)$ (Blumenthal
\etal, 1992; Gorski, 1992).

For a Harrison-Zel'dovich spectrum ($n=1$), consistent with the COBE
observations and the South Pole anisotropy upper-limit, we find that
$\delta M/M_{L} $ is $\approx 1.2$ at the scale of 5000 \kms, and thus
such voids are not expected to be typical.  As can be seen from the
figure, the typical diameter of volume-filling voids should be
$\approx$ 3500 \kms.  Voids as large as 5000 \kms could still be
formed from $2 \sigma$ fluctuations and they would represent the
high-end tail of the distribution of void sizes.

{}From figure 1 we can also see that the tilted CDM model considered by
Cen \etl (1992) ($n = 0.7$) can  produce gravitationally only small
voids, with diameters of about 2500 km/s. Voids with diameters of 5000
km/s would be very rare representing $5 \sigma$ events, in apparent
contradiction with the observations. This is to be expected
since this model was proposed to reconcile a high-bias CDM model with
the COBE results.  We remind the reader that the standard CDM model
($n=1$ on large scales) normalized to COBE leads to an unbiased galaxy
distribution with \dmm $\ap$ 1 at 800 \kms, in which case voids should
have small diameters, typically $\ll 2500 \kms$.

If the perturbations crossing the horizon are pure density fluctuations
then more power is needed to form voids. The required initial
amplitude is 4.5 and for a Harrison-Zel'dovich spectrum only $4
\sigma$ fluctuations will reach shell crossing on the 5000 \kms scale.
Clearly this is too rare to yield enough seeds to form the apparent
close-packed system of 5000 \kms voids.  The discrepancy that we find
is far larger than the uncertainty introduced from the fit of the
observed COBE data to the analytical curves.  In
this case, only voids with diameters of about $2500 \kms$ could grow
gravitationally.  We also note that for $n < 1$ only very small voids,
with diameters of about 1800 km/s, would form.

Of course, some  of the difficulties of forming large voids could
be alleviated if we consider the possibility that voids
are essentially formed when the underdensity is of the order of
30\%. In this case the required amplitudes are decreased by a
factor of 1.5 and voids with diameters of  4200 \kms can form
from 1$\sigma$ fluctuations for \dcrit=2.7, (3500 \kms for \dcrit=4.5)
for a Harrison-Zel'dovich spectrum.
We note, however, that in this case we would not expect
them to fill in the volume.  We also expect
the voids to have shallower profiles and
less sharp boundaries. Unfortunately, the available data is far
from adequate to address these details.

Alternatively, if $1.5 \sigma$ fluctuations are sufficient to produce a
volume filled Universe, (or if the voids are not volume filling) this
will also reduce by a similar factor the required amplitude for
formation of voids and 4200 \kms voids could form  with a $n=1$
spectrum.  However, we would have to relax both assumptions -- i.e. to
require that the observed voids are shallower and that they are
produced by $1.5 \sigma$ events -- to explain the
gravitational formation of 5000 \kms voids with such a spectrum.

\bigskip
\noindent
{\bf VI. DISCUSSION}
\bigskip
The new measurements of CMBR temperature anisotropy, especially as
measured by COBE, have offered the unique opportunity to normalize
the primordial power-spectrum of the matter distribution.
In this paper we have investigated how the properties
of voids observed in the galaxy distribution are constrained by the
CMBR anisotropy measurements.

The basic assumption of our model is that the ``voidy'' nature of the
galaxy distribution is a natural consequence of negative rms amplitude
fluctuations which grow in size during the evolution of the universe
producing an essentially void-filled distribution. Although the
details of our model are uncertain and the information on voids is
still sketchy, our formalism can be used to quantitatively assess the
likelihood of a given model of structure formation to generate voids.
The underlying assumption of our approach is that voids grow
gravitationally, that light traces matter at large scales and
therefore the large voids in the galaxy distribution correspond to
real voids in the matter distribution.  One might resort to biasing
and suggest that voids in the galaxies do not correspond to voids in
the matter. However, this would require a galaxy density which is not
proportional to the matter density on the scale of voids.  The
standard biasing schemes do not achieve this.  These schemes can
amplify, in the galaxy distribution, small amplitude perturbations in
the dark matter. However, for biasing to create a large void which
does not follow the dark matter distribution, it will also have to
erase structures corresponding to large amplitude dark matter
perturbations within the void.

Another basic assumption in our analysis is that the universe
is flat.  The dominant modification in an open Universe is the
increase in $R_h$ (which is proportional to $\Omega^{-1}$). This
results in a much larger separation between the scale of COBE ($\ge
10^o$) and the scale of voids ($5000 km/sec$). Thus the discrepancy
that we have pointed out between the $n=1$ power spectrum, the CMBR
observations and the appearance of voids may disappear in an open
universe.  The standard inflationary paradigm can thus be preserved
at the cost of introducing a non-zero cosmological constant.

Under the above general assumptions, our main conclusions
can be summarized as follows:

1) With our standard assumptions that \dcrit=2.7
and that $1 \sigma$ fluctuations produce the observed voids
(corresponding to a growing mode initial fluctuation and to
voids with an underdensity of 20\%)
we find:

\item{(a)} A power law of $n=1.25$ is required to produced a 5000 \kms void
filled Universe. This power spectrum gives $C$, on the scale of
$1^o$ slightly above the
limits of Gaier \etl (1992), but in agreement with the results
of Meyer (1993).

\item{(b)} A Harrison-Zel'dovich (n=1) spectrum can produce gravitationally,
only 3500 \kms voids.  Larger voids  can still be formed but are not
expected to be typical.

2) If we relax either one of these requirements (underdensity for voids
is 30\% or voids are produced by $1.5 \sigma$ fluctuations) the
required rms amplitude decreases by 1.5 and 5000 \kms voids can be
produced with $n=1.1$, while $n=1$ leads to 4200 \kms voids.

3) If we relax both assumptions we find that $n=1$ power spectrum
can produce 5000 \kms voids.

4) The ``common'' existence of voids on scales ranging from 2500 to
5000 \kms requires, if they form gravitationally, more power in the
range 800 to 1500 \kms than predicted from  the standard unbiased CDM
model.  This does not mean that occasionally voids of this size or
even larger cannot be produced by this model,
as demonstrated by some {\it N}--body
simulations.  However, they cannot be as common as those observed in the
galaxy distribution, in which they seem to be volume filling.  We can
also rule out gravitational formation of voids in the tilted CDM model
with $n = 0.7$, since for this model there is even less power on large
scales than in standard CDM.

5) The COBE data alone constrains the largest possible void to
about 6000 \kms in diameter.  If we also consider the limits imposed at
the $1^o$ scale then the size of the largest void is $<$~5000\kms.
Voids as large as those suggested by Broadhurst \etl (1991) cannot
form gravitationally and they cannot correspond to voids in the matter
distribution.

Based on our model, we believe that as long as voids grow
gravitationally, the existence and abundance of voids with sizes of
about 5000 \kms, as suggested by several redshift surveys, require
considerable power on scales larger than about 800\kms. This implies
that non-linear effects should be important in the evolution of
perturbations on these scales.  This contrasts with the conventional
idea that scales beyond about 8 \h1 Mpc are still in the linear
regime.

If we assume that rich cluster form from positive density fluctuations
with the same initial amplitude as the negative perturbtions that form
voids, we expect that in our model there will be a relationship
between the abundance of voids and rich clusters.  For a void-filled
universe with a scale of 5000 \kms the abundance of dark matter
clusters with masses of $5 \times 10^{15} M_{\odot}$ (corresponding
roughly to galaxy clusters of $5 \times 10^{14} M_{\odot}$) will be of
the order of one per $(50 h^{-1} Mpc)^3$. This should be
compared with an Abel richness 1 cluster density of one per
$(55h^{-1}Mpc)^3$ and with the density of Abell clusters of richness
2, one per $(95h^{-1}Mpc)^3$ (Bahcall and Cen, 1992).  It seems that
our model predicts more rich clusters than observed. However, given
all the uncertainties involved in the model and the data available for
clusters we cannot discard the possibility of non-linearity on the
basis of the existing data on cluster abundance.  Specifically, if
voids form from $1.5 \sigma$ fluctuations the symmetry between
positive and negative fluctuations is broken, reducing the predicted
cluster density.

We would argue that the mounting evidence for the common existence of
large voids points out the need to pursue more evolved N--body
simulations, allowing for larger values of $\sigma_8$ (e.g. Couchman
\& Carlberg 1992).  These simulations would also be important to
confirm if our model for the formation of voids from primordial
underdense regions is correct.  We note that for $n=1$, $\sigma_8$ can
be as high as 4.5 and still be consistent with the newly established
limits of the UCSB experiment. Turning the argument around, if the
value of $\Delta T/T$ at the few degree scales is of the same order as
the current upper-limits this will definitely establish that the
matter distribution is much more evolved than originally thought.  The
challenge will then be to account for the properties of galaxy
clustering in small scales and the relationship between the galaxy and
the matter distribution.

Although we have only considered generic power-law spectra, our
approach can also be used to predict the size of typical voids for any
specific model of structure formation.  It would also be interesting
to examine  how the power requirement for void formation
compares to that necessary to account for the bulk motion on
comparable scales.  It is interesting that the latter seems to be the
most critical test for all alternative models proposed to replace
standard CDM.  As in the case of voids, the observed bulk motions seem
to require extra power on large--scales, which as discussed by Gorski
(1992) conflicts with the upper limit imposed the UCSB experiment.  It
should be pointed out that since the existence of large voids seems to
indicate that non-linear effects may be important on scales as large
as 5000 \kms, the underlying assumptions used to compare the
theoretical predictions for the amplitude of bulk motions with the
observations may not be valid. This demonstrates that CMBR
measurements, bulk motion data and size of voids may offer
complementary information on the nature of the primordial
power--spectrum and are important independent tests which may help to
discriminate amongst competing models of structure formation.

\noindent {\bf Acknowledgments}

We thank  S. Meyer for providing us with his data and
Rien van Weygaert for providing us with his Voronoi tesselation code.
LdC and TP would like to thank the CFA for its hospitality during the
course of this research.  This work was supported in
part by NSF grant PHY9024920 (GRB), by a Center for Astrophysics
Fellowship (DSG), and by the Smithsonian Institution's Visitor Fund
(LdC,TP).

\bigskip
\line{\bf References:\hfil} \nobreak
\baselineskip=15pt
\ppp
Bahcall, N. \& Cen, R., (1992) Ap. J. Lett. {\bf 398}, L81.
\ppp
Bertschinger, E., \etal, (1990) Ap. J., {\bf 364}, 370.
\ppp
Blumenthal, G., R., Da Costa, L., Goldwirth, D., S., Lecar, M. \&
Piran, T., (1992),  Ap. J.. {\bf 388}, 234.
\ppp
Broadhurst, T.\ J., Ellis, R.\ S.,  Koo, D.\ C., \& Szalay, A.\ S.\ (1990)
Nature, {\bf 343}, 726.
\ppp
Cen, R. \etal, (1992) Ap. J. Lett.  {\bf 399}, L11.
\ppp
Cen R. \& Ostriker, J., P., (1992a) Ap. J. {\bf 393}, 22.
\ppp
Cen R. \& Ostriker, J., P., (1992b) Ap. J. Lett. {\bf 399}, L113.
\ppp
Couchman, H.M.P. \& Carlberg, R.G., (1992), Ap. J., {\bf 389}, 453.
\ppp
Courteau S. (1992) USC PhD thesis.
\ppp
da Costa, L. N., Pellegrini, P. S., Sargent, W. L. W., Tonry, J., Davis, M.,
Meiksin, A. Latham, D. W., Menzies, J. W., \& Coulson, I. A.
(1988) Ap. J. {\bf 327}, 544.
\ppp
da Costa, L. N. (1991) in {\it The Distribution of Matter in the Universe}
ed. D. Gerbal \& G. A. Mamon (Meudon:Observatoire de Paris).
\ppp
Davis, M. Summers, F. J.  \& Schlegel D. (1992) Nature {\bf 359}, 393.
\ppp
Dey, A., Strauss, M. A., \& Huchra, J., (1990) AP, {\bf 99}, 463.
\ppp
Dubinski, J. da Costa, L. N., Goldwirth, D. S., Lecar, M. \& Piran, T.
(1993) Ap. J., in press
\ppp
Efstathiou, G. \etal, MNRAS, {\bf 247}, 10p.
\ppp
Efstathiou, G. Bond, J.R., \& White D. M.,  MNRAS, {\bf 258}, 1p.
\ppp
Gaier, T. \etal, (1992) Ap. J. Lett. {\bf 398}, L1.
\ppp
Geller, M.\ J., \&  Huchra, J.~P.~(1989) Science {\bf 246}, 897.
\ppp
Gorski, K. (1991), Ap. J. Lett. {\bf 370}, L5.
\ppp
Gorski, K. (1992), Ap. J. Lett. {\bf 398}, L5.
\ppp
Gould, A., (1993) Ap. J. Lett. {\bf 403}, L51.
\ppp
Kirshner, R.\ P., Oemler, A. Jr., Schechter, P.\ L., \&
Shectman, S.\ A.\ (1981) Ap. J. Lett. {\bf 248}, L57.
\ppp
Kirshner, R. P., Oemler, A. Schechter, P. L., \& Shectman, S. A.
(1983) AJ, 88, 1285.
\ppp
Kirshner, R. P., (1993) Private communication.
\ppp
Kofman , L.A, Gnedin, N. Y. \& Bahcall, N.A. (1993) Ap. J. submitted.
\ppp
De Lapparent, V., Geller, M.\ J., \& Huchra, J.\ P.\ (1986)
Ap. J. Lett. {\bf 302}, L1.
\ppp
Klypin,A. , Holtzman, J. Primack, J. \& Regos, E. (1992) preprint.
\ppp
Maddox, S., J., \etal, (1990a) MNRAS,  {\bf 242}, 43p.
\ppp
Maddox, S., J., \etal, (1990b) MNRAS,  {\bf 242}, 4433.
\ppp
Mart\'{\i}nez-Gonz\'alez, E., \& Sanz, J.\ L.\  (1989) Ap. J. {\bf 347}, 11.
\ppp
Meyer, S., \ S., (1993) private communication.
\ppp
Sachs, R.\ K.\  \&  Wolfe, A.\ M.\  (1967), Ap. J. {\bf 147}, 73.
\ppp
Saunders, W., Rowan-Robinson, M., \&Lawrence, A., (1992) MNRAS,
{\bf 258}, 134.
\ppp
Schectman, S. A., Schechter, P. L., Oemler, A. A., Tucker, D.,
Kirshner, R. P. \& Lin, H., (1992) to appear in Clusters and Superclusters
of Galaxies, ed A. C. Fabian.
\ppp
Smoot, G.\ F.\ \etl (1992) Ap. J. Lett. {\bf 396}, L1.
\ppp
Van de Weygaert, R., (1991) Ph.D thesis Leiden University.
\ppp
Van de Weygaert, R., \& Van Kampen, (1993) MNRAS, in press.
\ppp
Vogeley, M.S., Park, C., Geller, M.J. \& Huchra, J.P., 1992, Ap. J.,
{\bf 391}, L5.
\ppp
Wright, E.\ L.\ \etl (1992), Ap. J. Lett. {\bf 396}, L13.

\vfill\eject
\bigskip\bigskip
\baselineskip=18pt
\line{\bf Figure Captions\hfil}\nobreak
\ppp
{\bf Fig.\ 1:}
Predicted \dmm vs the void diameter (measured in km/sec)
for various power laws spectrum, whose amplitude fits COBE.

\bigskip\bigskip
{\bf Table I}

$$\vbox{\tabskip=0pt \offinterlineskip
\halign to\hsize{\strut#
& \vrule#\tabskip=1em plus2em&\hfil#& \vrule#& \hfil#& \vrule#
&\hfil#& \vrule#& \hfil#& \vrule#& \hfil#&\vrule#
&\hfil#&\vrule#
&\hfil#&\vrule#
\tabskip=0pt\cr \noalign{\hrule}
&&  $n$ &&  $\sqrt{\VEV{{ \tilde C_{obs} \over T^2 \tilde F}}} $
&&  $\sqrt{{C_{2~(obs)}^2\over  T^2 F_{2}}} $
&&  ${\delta M \over  M_L}  \ ^1 $
&& $v  \ ^2 $  && $\sqrt{C_{1.2}(0)}  \ ^3$ && $\sqrt{C_{3.8}(0)}  \ ^4$
&\cr \noalign{\hrule}
&&   .50  && $1.3 \cdot 10^{-5}$  && $.98\cdot 10^{-5} $
&& 0.34  \hfil  && 1500  &&
$1.3\cdot 10^{-5}$  &&$1.2\cdot 10^{-5}$
&\cr \noalign{\hrule}
&&   .75  && $.1 \cdot 10^{-5}$  && $.85\cdot 10^{-5}$
&& 0.63 \hfil  && 2400   &&
$1.4\cdot 10^{-5}$  &&$1.3\cdot 10^{-5}$
&\cr \noalign{\hrule}
&&  1.00  && $.89 \cdot 10^{-5}$ && $.74\cdot 10^{-5}$
&& 1.2 \hfil   && 3400    &&
$1.6\cdot 10^{-5}$  &&$1.4\cdot 10^{-5}$
&\cr \noalign{\hrule}
&&  1.25  && $.74 \cdot 10^{-5}$ && $.64\cdot 10^{-5}$
&& 2.4 \hfil   && 4800   &&
$1.9\cdot 10^{-5}$  &&$1.6\cdot 10^{-5}$
&\cr \noalign{\hrule}
&&  1.50  && $.65 \cdot 10^{-5}$ && $.54\cdot 10^{-5}$
&& 4.7 \hfil   &&  6500  &&
$2.4 \cdot 10^{-5}$  &&$2.\cdot 10^{-5}$
&\cr \noalign{\hrule}
\noalign{\smallskip} \hfil\cr}}$$

1) $\delta M/M_L $ on a scale that corresponds to
voids with a diameter of $v=5000 km/sec $ today.

2) Diameter of the voids (using $\delta M/M_L = 2.7$) in $km/sec$.

3) $\sqrt{C(0)}$ convolved with a FWHM beam of $1.2^0$. This sould be
compared with the $2 \sigma$ upper limit of Gaier \etl (1992) of $1.4 \cdot
10^{-5}$.

4) $\sqrt{C(0)}$ convolved with a FWHM beam of $3.8^0$. This sould be
compared with the detection of Meyer  (1993) of $1.4 \pm .5 \cdot
10^{-5}$.

\bye